\documentclass[preprint,12pt]{elsarticle}

\usepackage{amssymb}

\journal{Physica A}

\begin{document}

\begin{frontmatter}



\title{Cultural evolution and personalization}


\author[inst1]{Ning Xi}
\author[inst2]{Zi-Ke Zhang}
\author[inst1,inst3]{Yi-Cheng Zhang}

\address[inst1]{Business School, University of Shanghai for Science and Technology, Shanghai, 200093, P. R. China}
\address[inst2]{Institute for Information Economy, Hangzhou Normal University - Hangzhou 310036, P. R. China}
\address[inst3]{Department of Physics, University of Fribourg,Chemin du Mus\'ee 3, 1700 Fribourg, Switzerland}



\begin{abstract}
In social sciences, there is currently no consensus on the mechanism for cultural evolution. The evolution of first names of newborn babies offers a remarkable example for the researches in the field. Here we perform statistical analyses on over $100$ years of data in the United States. We focus in particular on how the frequency-rank distribution and inequality of baby names change over time. We propose a stochastic model where name choice is determined by personalized preference and social influence. Remarkably, variations on the strength of personalized preference can account satisfactorily for the observed empirical features. Therefore, we claim that personalization drives cultural evolution, at least in the example of baby names.

\end{abstract}


\end{frontmatter}


\section{Introduction}

Cultural evolution is a dynamical process that cultural traits change over time due to species' fitness to social and natural environment. On one hand,

 and thus it can be quantitatively described by the distribution of cultural traits.

Remarkably, at all times evolutionary process exhibits the similar statistical
character that a relatively small number of traits are very popular, however, the majority barely gets any attention at all.
In the past few decades, a wide range of studies have been carried out in an attempt to uncover the mechanism generating such inequality.
One explanation is given by Rosen and MacDonald~\cite{Rosen,MacDonald}. They suggest that the inequality is caused by the differential
quality of cultural traits and can be reproduced by convexity of the mapping from quality to popularity. An alternative explanation is
provided by Adler~\cite{Adler}. He argues that individuals' decisions are influenced by the behavior of others, which leads to the inequality.

In order to test the empirical validity of the theories, Hamlen examined the relationship between voice quality and record sales in the
popular music industry~\cite{Hamlen}. Empirical results show that the estimated elasticity of record sales to voice quality is less than one,
which repudiates the explanation of Rosen and MacDonald. Afterward, Chung and others studied the role of social influence in success with the
data from the Gold-Record Awards~\cite{Chung}. They used the number of gold-records as the measure of success and found that the stochastic
model incorporating social influence can explain the observed inequality in the empirical data excellently. Recently, Salganik and others
investigated social influence in cultural markets by a well-designed web-based experiment in which participants may download previously unknown
songs either with or without knowledge of previous participants' choices~\cite{Salganik}. Comparative experiment shows that both the convex
mapping from quality to popularity and social influence play the vital role in the emergence of inequality.

Besides inequality among cultural traits, how the inequality evolves is also a significant topic in researches on cultural evolution. However,
so far we almost know nothing about it, partly because of lack of suitable data. Luckily, the evolution of first names of newborn babies offers a remarkable example for the researches. In the paper, we perform statistical analyses on over $100$ years of data in the United States to investigate the following: (1) The frequency-rank distribution and its evolution; (2) The evolution of inequality; and (3) The property of temporal autocorrelation. Guided by the empirical results, we propose a stochastic model where name choice is determined by personalized preference and social influence. We show that the simple model can reproduce the observed empirical features very well.

\section{Data Analysis}

The data on first names are taken from US Social Security Administration, and contain the top $1000$ boys' and girls' names every year from $1880$ to $2010$. All names are from Social Security card applications for births that occurred in the United States after $1879$. All data are from a 100\% sample of the records on Social Security card applications as of the end of February $2011$.

Firstly, we study the distribution of baby names and its evolution. As shown in Fig.$1$ (a), the frequency-rank distribution of baby names follows the two-regime power law where the first power law decay has a smaller exponent than the second one. The law was also found in the studies on the frequency of words~\cite{Cancho}. Then we compare the distributions in different years and find that both the exponents in two regimes decline over time. Fig.$1$ (b) graphically illustrates the evolution of the distribution, taking four distributions for instance.

\begin{figure}
\centering
\includegraphics[width=0.5\textwidth]{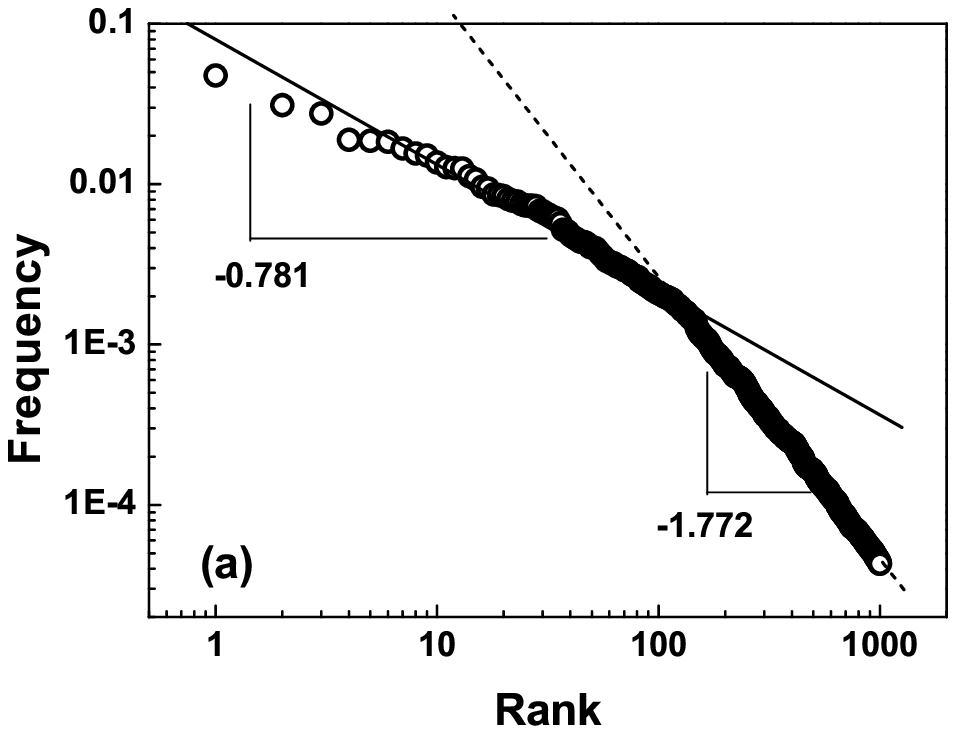}%
\includegraphics[width=0.5\textwidth]{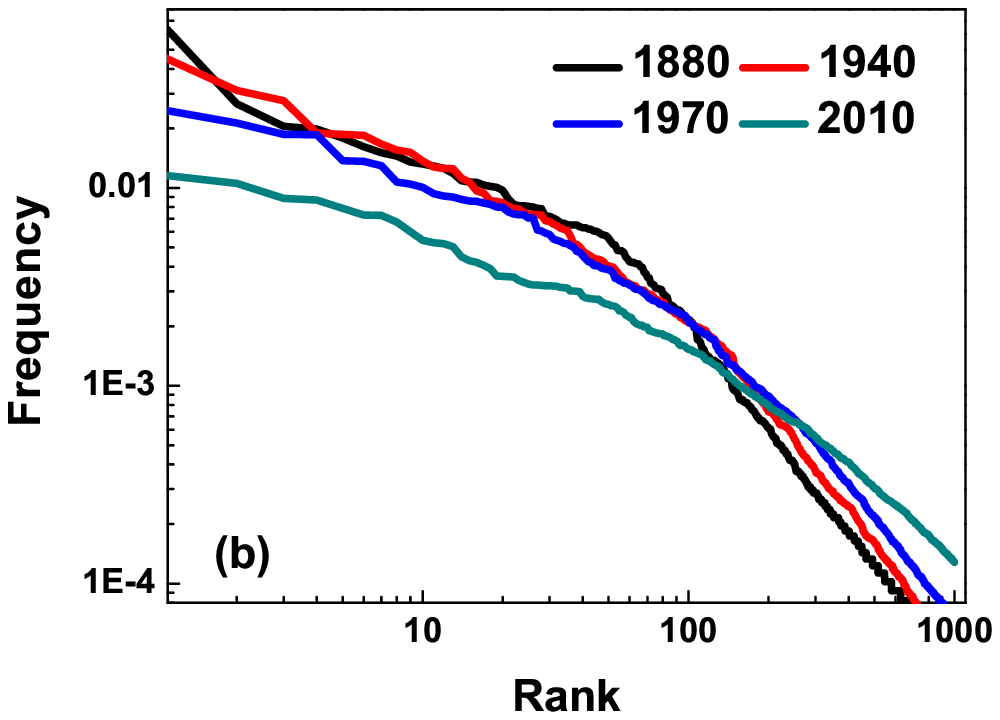}
\caption{The frequency-rank distribution of baby girl names. (a) shows the distribution in $1940$, where the first and second power law decays have exponents $0.781\pm0.010$ and $1.772\pm0.002$, respectively. (b) shows the distributions in $1880$, $1940$, $1970$ and $2010$. By comparison, one can find that both exponents in two regimes decrease over time. Baby boy names have similar statistical features.}
\end{figure}

Secondly, we focus on the evolution of inequality. We use Simpson's index to measure the inequality among baby names. Simpson's index is defined as the probability of any two individuals drawn at random from newborn babies in a year choosing the same first name, and is expressed as follows~\cite{Magurran},
\begin{equation}\label{simpson}
    I=\sum^n_{i=1}{p_i^2},
\end{equation}
where $p_i$ denotes the frequency of baby name $i$, and $n$ is the number of first names in a year. It ranges from $1/n$ (complete equality) to one (maximum inequality). Simpson's index is heavily weighted towards names with large frequency, while being less sensitive to the lack of names with small frequency. Our data omit the names outside the top $1000$, and thus Simpson's index is the most suitable measure of the inequality for our studies. We calculate Simpson's index for each year, and the results are shown in Fig.$2$. Inequality, in the main, declines over time.

\begin{figure}
  \centering
  \includegraphics[width=0.7\textwidth]{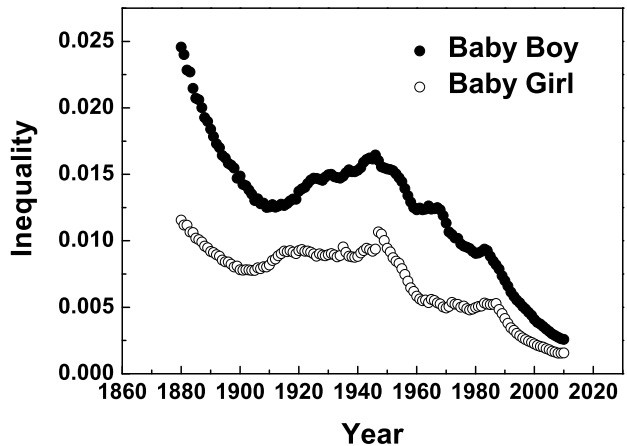}
  \caption{The evolution of inequality. Inequality, in the main, declines over time.}
\end{figure}

Thirdly, we study the property of temporal autocorrelation of the data. Consider any two years $t$ and $t+\triangle{t}$. The same baby names are picked up from the data in the two years, and their used times in the two years are expressed as two vectors $y_t$ and $y_{t+\triangle{t}}$, respectively. Correlation is defined as Pearson's correlation coefficient between $y_t$ and $y_{t+\triangle{t}}$, which is computed by the covariance of the two vectors divided by the product of their standard deviations. The formula is expressed as follows,
\begin{equation}\label{correlation}
    C(t,\triangle{t})=\frac{E[(y_t-\mu_t)(y_{t+\triangle{t}}-\mu_{t+\triangle{t}})]}{\sigma_t\sigma_{t+\triangle{t}}}
\end{equation}
The empirical results are shown in Fig.$3$. For any given value of $\triangle{t}$, the correlation $C(t,\triangle{t})$ drops with time $t$.
\begin{figure}
  \centering
  \includegraphics[width=0.7\textwidth]{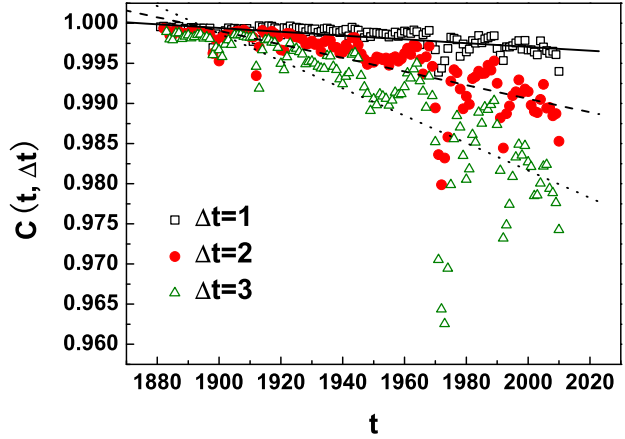}
  \caption{The correlation functions of the data on baby names. The linear fits to the data show that the correlation $C(t,\triangle{t})$ drops with time $t$. In the figure, we only take the three specific values of $\triangle{t}$ for instance.}
\end{figure}

\section{The Stochastic Model of Cultural Evolution}

To gain a deeper insight into cultural evolution, we propose a stochastic model with the assumption on the individual's decision making to reproduce all the observed empirical features. In the artificial society, there are $N$ names to choose from. Time is discrete. At each time step, $B$ new individuals are born and choose names according to their evaluations for these names. Obviously, the individual's evaluation is based on personalized preference. The more an individual likes a name, the more likely he chooses it. Besides personalized preference, the individual's evaluation is also socially influenced, which can be seen from the fact that any one tends to choose the name that he likes and that others also think well of. Based on this, we give a formulas, by which individual $i$'s evaluation for name $j$ at time $t$ can be computed, as follows,
\begin{equation}\label{evaluation}
    p_{ijt}=\omega\frac{Q_{ijt}}{\displaystyle\sum^{N}_{k=1}Q_{ikt}}+(1-\omega)\frac{\displaystyle\sum^{t-1}_{l=t-m}a_{jl}}{\displaystyle\sum^{N}_{k=1}\sum^{t-1}_{l=t-m}a_{kl}},
\end{equation}
where $Q_{ijt}$ denotes individual $i$'s preference to name $j$ at time $t$, and $a_{kl}$ is the used times of name $k$ at time $l$. In reality, the effect of the used times on the individual's evaluation is far from being uniform in time~\cite{Zanette,Dorogovtsev,Cattuto}. Thus, in Eq.($3$), only the used times in recent $m$ time steps are considered. In terms of our model, $\sum^{N}_{k=1}\sum^{t-1}_{l=t-m}a_{kl}$ is equal to $mB$. $\omega$ is the weight which ranges from $0$ to $1$. When $\omega$ is high, the evaluation process is considered to be more personalized. Similar equations were used in the studies on other issues~\cite{Gould}. Here, for simplicity, we assume that all the names are identical to all the individuals at any time step. Thus, Eq.($3$) changes to the following form
\begin{equation}\label{simplicity}
    p_{ijt}=\omega\frac{1}{N}+(1-\omega)\frac{\displaystyle\sum^{t-1}_{l=t-m}a_{jl}}{mB}.
\end{equation}
Individuals choose names with the probability proportional to their evaluations for names computed by Eq.($4$). From this equation, we can also infer that temporal autocorrelation we define decreases with the increase in $\omega$. Recall the empirical study on temporal autocorrelation. The observed decline of temporal autocorrelation may suggest the increase in the strength of personalized preference $\omega$.

We ran computer simulations of the stochastic model and collected the used times of each name in $R$ time steps after reaching the steady state. As shown in Fig.$4$, the frequency-rank distribution of baby names follows two-regime power law, consistent with the empirical result. At present,
a vital issue to be solved is what drives the process of cultural evolution. Luckily, the empirical study on temporal autocorrelation has given us a key hint that the strength of personalized preference becomes strong with the evolution. We checked whether the increase in the strength of personalized preference generates the evolution by computer simulations with various values of $\omega$. The results are shown in Fig.$5$. It can be found that with the increase in the strength of personalized preference, both the exponents in two regimes of the frequency-rank distribution decline and the inequality also decreases, extremely similar to the empirical observations. Thus, we assert that it is personalization to drive the process of cultural evolution, at least in the example of baby names.

\begin{figure}
  \centering
  \includegraphics[width=0.7\textwidth]{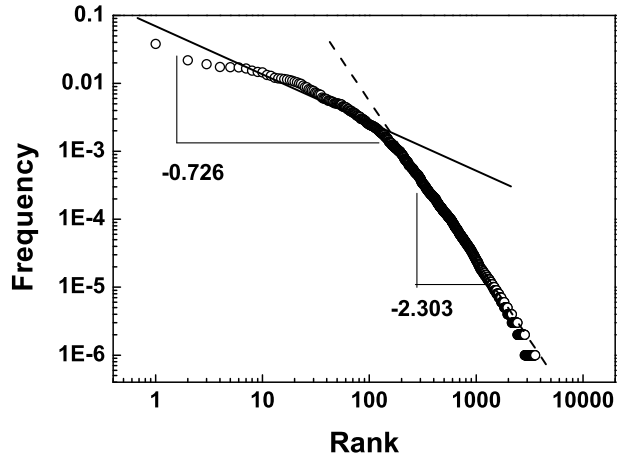}
  \caption{The frequency-rank distribution of baby names, resulting from the run of the computer simulation with $N=6000$, $B=250$, $m=100$, $\omega=0.005$ and $R=4000$. The distribution follows two-regime power law, where the first and second power law decays have exponents $0.726\pm0.015$ and $2.303\pm0.003$, respectively.}
\end{figure}

\begin{figure}
\centering
\includegraphics[width=0.5\textwidth]{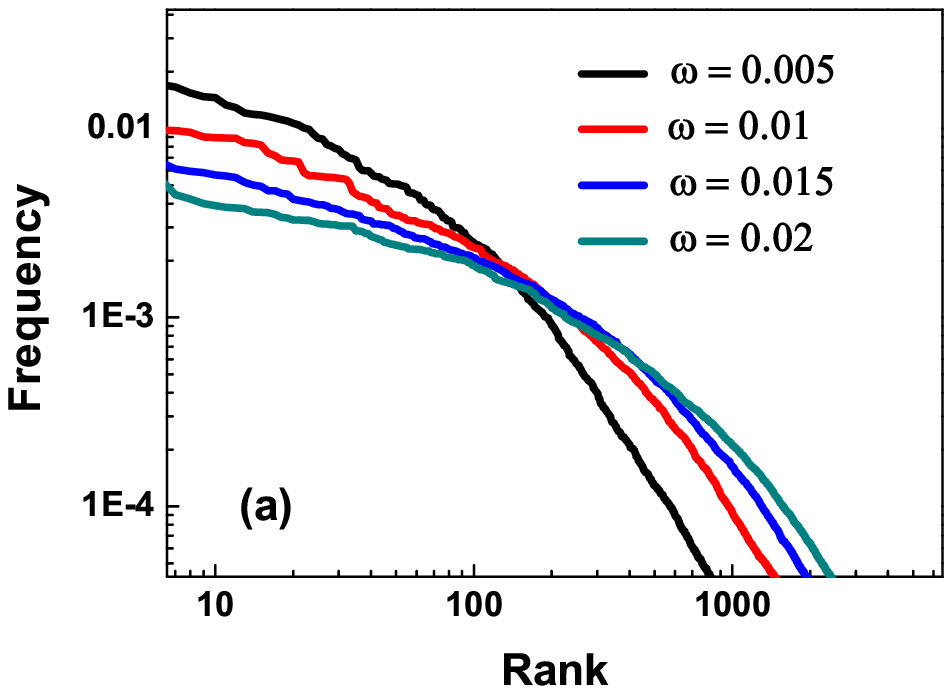}%
\includegraphics[width=0.5\textwidth]{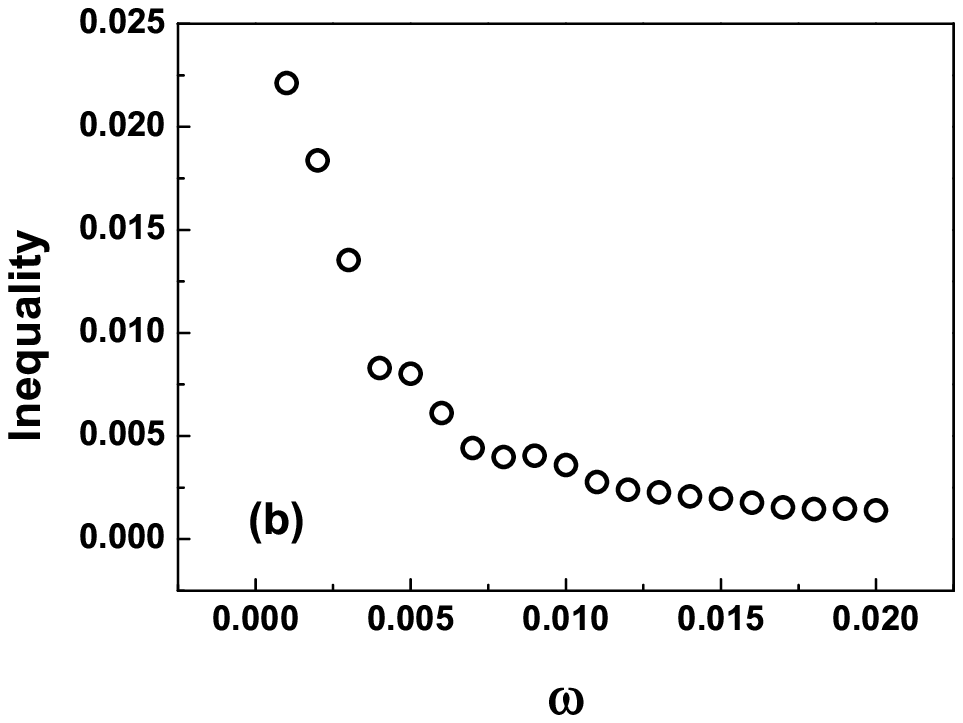}
\caption{The evolution of baby names shown by computer simulations. (a) shows the evolution of the frequency-rank distribution, resulting from the simulations with $N=6000$, $B=250$, $m=100$, $R=4000$ and: $\omega=0.005$ (black); $\omega=0.01$ (red); $\omega=0.015$ (blue); $\omega=0.02$ (green). Both the exponents in two regimes of the frequency-rank distribution decline with increasing $\omega$. (b) shows the change of the inequality with $\omega$. When $\omega$ increases, the inequality decreases.}
\end{figure}

\section{Conclusion}

In this paper, we take baby names for instance to investigate the process of cultural evolution, both empirically and theoretically. In the empirical studies, firstly we find that the frequency-rank distribution of baby names follows two-regime power law and both the exponents in two regimes decrease over time. Secondly, we use Simpson's index to measure the inequality among baby names and reveal the decline of inequality. Thirdly, we define the temporal autocorrelation function and indicate its decaying with time. To uncover the driving force of cultural evolution, we propose a simple stochastic model where the individual's decision making is determined by personalized preference and social influence. Computer simulations show that the increase in the strength of personalized preference can produce the patterns quite similar to the empirical observations. Based on this, we claim that personalization drives cultural evolution.



\end{document}